\documentclass[12pt,english,a4paper]{article}
\usepackage{natbib}
\usepackage{rotating,placeins}
\usepackage{babel}
\usepackage{geometry}
\usepackage{fancyhdr}
\usepackage{lscape,epsfig,amssymb,amsmath}
\usepackage[]{subfigure}
\usepackage{rotating,placeins}
\usepackage{lscape}
\usepackage{setspace}
\newcommand{\UR}{\overline{u'_i u'_j}}
\newcommand{\dr}{\mathrm{d}}
\newcommand{\Dr}{\mathrm{D}}
\newcommand{\er}{\mathrm{e}}
\newcommand{\Sec}[1]{Sec~\ref{sec:#1}}
\newcommand{\Fig}[1]{Fig~\ref{fig:#1}}
\newcommand{\Eq}[1]{Eq~(\ref{eq:#1})}
\newcommand{\Dp}[2]{\frac{\partial #1}{\partial #2}}

\title{Approximate Stokes Drift Profiles in Deep Water}
\author{{\O}yvind Breivik
  \footnote{Final version to be published in \emph{J Phys Oceanogr}, doi:10.1175/JPO-D-14-0020.1 (in press)}
  \thanks{Corresponding author. E-mail:
          \texttt{oyvind.breivik@ecmwf.int}. 
          ORCID Author ID: \texttt{0000-0002-2900-8458}}
  \thanks{ECMWF, Shinfield Park, Reading, RG2 9AX, UK.} 
  \and Peter A~E~M Janssen\footnotemark[3]
  \and Jean-Raymond Bidlot\footnotemark[3]}

\begin{document}
\maketitle

\abstract{
A deep-water approximation to the Stokes drift velocity profile is
explored as  an alternative to the monochromatic profile. The alternative
profile investigated relies on the same two quantities required for the
monochromatic profile, viz the Stokes transport and the surface Stokes
drift velocity. Comparisons with parametric spectra and profiles under
wave spectra from the ERA-Interim reanalysis and buoy observations reveal
much better agreement than the monochromatic profile even for complex sea
states.  That the profile gives a closer match and a more correct shear
has implications for ocean circulation models since the Coriolis-Stokes
force depends on the magnitude and direction of the Stokes drift profile
and Langmuir turbulence parameterizations depend sensitively on the shear
of the profile. The alternative profile comes at no added numerical cost
compared to the monochromatic profile.}

\section{Introduction}
\label{sec:intro}
With the inclusion of Langmuir turbulence\- (\citealt{sky95},
\citealt{mcwilliams97}, \citealt{thorpe04}, \citealt{ardhuin06},
\citealt{grant09} and \citealt{belcher12}) and Coriolis-Stokes forcing
(\citealt{hasselmann70}, \citealt{weber83}, \citealt{jenkins87b},
\citealt{mcwilliams99}, \citealt{janssen04}, \citealt{polton05} and
\citealt{janssen12}) in Eulerian ocean models it becomes important to
model the magnitude and the shear of the Stokes drift velocity correctly.
Stokes drift profiles are also needed when estimating the drift of partially
or entirely submerged objects (see \citealt{mcwilliams00}, \citealt{bre12b},
\citealt{rohrs12} and references by \citealt{bre13} for applications of
Stokes drift velocity estimates for particle and object drift). However,
computing the Stokes drift profile is expensive since it involves evaluating
an integral with the two-dimensional wave spectrum at every desired
vertical level. It is also often impractical or impossible since the full
two-dimensional (2-D) wave spectrum may not be available.  For this reason
it has been customary to replace the full Stokes drift velocity profile by
a monochromatic profile matched to the transport and the surface Stokes
velocity [see e.g.  \citet{sky95}, \citet{mcwilliams00}, \citet{car05},
\citet{polton05}, \citet{saetra07}, \citet{tamura12}].  This is problematic,
since it is clear that the shear under a broad spectrum is much stronger than
that of a monochromatic wave of intermediate wave\-number due to the presence
of short waves whose associated Stokes drift quickly vanishes with depth. At
the same time, the deep Stokes drift profile will be stronger than that of
a monochromatic wave since the low wave\-number components penetrate much
deeper. It is therefore of interest to investigate profiles that exhibit
stronger shear near the surface and a stronger deep drift.  Here we explore
an alternative approximate Stokes drift profile which will be compared
to the monochromatic profile.  The computation of the profile follows the
same procedure as when estimating a monochromatic profile. The alternative
profile has a lower mean-square error (MSE) than the monochromatic profile
for all spectra tested, as will be shown in detail in later sections. It has
a stronger shear in the upper part and does not tend to zero as rapidly as
the monochromatic profile in the deeper part.  This mimics the effect of a
broader spectrum where the low wave\-number components penetrate deeper than
the mean wave\-number component while the shorter waves (higher wave\-numbers)
only affect the the upper part of the water column.  The proposed profile has
the advantage of being robust, easy to implement and being computationally
inexpensive. Importantly, it relies on the same two integrated parameters
required to compute the monochromatic profile, namely the surface Stokes
drift velocity and the Stokes transport.  The proposed profile was recently
implemented (see \citealt{janssen13} and \citealt{bre13e}) in the European
Centre for Medium Range Weather Forecast's (ECMWF) implementation of the NEMO
ocean model [\cite{madec12}; the coupled forecast system and the coupling
between the wave model and the ocean model components are described by
\citet{janssen13} and \citet{mogensen12}].

This paper is organized as follows. In \Sec{profiles} we derive the analytical
expression for the monochromatic Stokes drift profile and the alternative
profile. In \Sec{paramspectra} we investigate how these two approximate
profiles compare for three well-known parametric spectra. \Sec{wamspectra}
examines the impact of a high-frequency spectral cut-off on the Stokes drift
profile and the Stokes transport. This has implications for the computation
of profiles from discretized spectra from numerical wave prediction models
\citep{has88,tol91,kom94,boo99,ris99,tolman02,jan04}. We investigate how well
the proposed profile fits the full profiles computed from two-dimensional wave
spectra from the ERA-Interim reanalysis \citep{dee11} in \Sec{erai}. Here
we also quantify how much waves beyond the high-frequency cut-off affect
the shear and the magnitude of the Stokes drift profile  (this was also
investigated by \citealt{rascle06}).  Furthermore, we investigate the impact
of approximating the Stokes transport direction by the more readily available
mean wave direction as well as approximating the magnitude of the Stokes
transport vector by the first order moment.  \Sec{ekofisk} investigates
profiles under observed wave spectra at Ekofisk in the North Sea. Finally,
in \Sec{conclusion} we present our recommendations for the computation of
approximate Stokes drift profiles.

\section{Approximate Stokes Drift Profiles}
\label{sec:profiles}
The Stokes drift profile in water of arbitrary depth was shown by
\citet{kenyon69} to relate in the case of linear waves to the wave variance
spectrum as
\begin{equation}
   \mathbf{v}_\mathrm{s}(z) = g\int\!\int_{-\infty}^{\infty} F(\mathbf{k}) \frac{\mathbf{k}}{\omega}
   \left[\frac{2k \cosh 2k(z+h)}{\sinh 2kh}\right] \, \dr\mathbf{k},
   \label{eq:ugeneral}
\end{equation}
where $k = |\mathbf{k}|$ is the magnitude of the wave\-number vector, $h$ is the
bottom depth (positive), $g$ the gravitational acceleration, $\omega = 2\pi
f$ the circular frequency and $z$ is the vertical co-ordinate (positive up).
To avoid confusion we use $v$ for Stokes drift velocities and $u$ for
Eulerian currents.  In the following we will only consider the deep-water
limit of the dispersion relation,
\begin{equation}
   \omega^2 = gk.
   \label{eq:disp}
\end{equation}
Then \Eq{ugeneral} simplifies to
\begin{equation}
   \mathbf{v}_\mathrm{s}(z) = \frac{2}{g}\int\!\int_{-\infty}^{\infty} \omega^3 
    \hat{\mathbf{k}} \er^{2kz} F(\mathbf{k}) \, \dr\mathbf{k},
   \label{eq:udeep}
\end{equation}
where $\hat{\mathbf{k}} = \mathbf{k}/k$ is the unit vector in the direction
of the wave component.

We now recast the east and north components of
the Stokes drift profile in frequency-direction $(f,\theta)$ co-ordinates as
\begin{equation}
   \mathbf{v}_\mathrm{s}(z) = \frac{16\pi^3}{g} \int_0^{2\pi}\! \int_0^{\infty}
   f^3 \hat{\mathbf{k}} \er^{2kz}
    F(f,\theta) \, \dr f \dr\theta,
   \label{eq:uvfth}
\end{equation}
where $\theta$ is measured clockwise from north (going to).  The Stokes
transport $\mathbf{V}_\mathrm{s} = \int_{-\infty}^0 \mathbf{v}_\mathrm{s}(z)
\, \dr z$ becomes in the deep-water limit
\begin{equation}
   \mathbf{V}_\mathrm{s} = 2\pi \int_0^{2\pi}\!\int_0^{\infty} f \hat{\mathbf{k}} F(f,\theta) 
    \, \dr f \dr\theta.
  \label{eq:UV}
\end{equation}
The integrand here is the first-order moment of the wave spectrum, $m_1$,
weighted by the unit vector $\hat{\mathbf{k}}$ of the wave component, with
the $n$-th order moment of the 2-D spectrum defined as
\begin{equation}
   m_n = \int_0^{2\pi}\!\int_0^{\infty} f^n F(f,\theta) \, \dr f \dr\theta.
  \label{eq:mn}
\end{equation}

Estimating the full profile from \Eq{uvfth} can be a costly operation even
when a modeled or observed wave spectrum is available.  When a wave spectrum is
not available the Stokes drift profile must be approximated from the transport
(\ref{eq:UV}) and the surface Stokes drift velocity.  It is therefore common
to approximate \Eq{uvfth} by the exponential profile of a monochromatic wave
[see eg \cite{sky95,mcwilliams00,car05,polton05,saetra07,tamura12}]
\begin{equation}
   v_\mathrm{m} = v_0 \er^{2k_\mathrm{m}z}.
   \label{eq:uvmono}
\end{equation}
To ensure that the surface Stokes drift and the total transport of the
monochromatic wave in \Eq{uvmono} agree with the values for the full spectrum,
Eqs~(\ref{eq:uvfth})-(\ref{eq:UV}), the wave\-number must be determined by
\begin{equation}
   k_\mathrm{m} = \frac{v_0}{2{V}_\mathrm{s}}.
   \label{eq:kmono}
\end{equation}
A monochromatic profile will have a weaker vertical gradient than
the profile under a full spectrum near the surface whereas it tends
too quickly to zero deeper down.  The behavior of the profile under a
full spectrum is most readily investigated by considering the Phillips spectrum
\citep{phillips58,phi85,jan04}, applicable to the equilibrium range of the
spectrum of wind-generated waves above the spectral peak,
\begin{equation}
   F_\mathrm{P} = \left\{ \begin{array}{lr}
             \alpha_\mathrm{P} g^2 \omega^{-5}, & \omega > \omega_\mathrm{p} \\
             0,                                 & \omega \leq \omega_\mathrm{p} 
                          \end{array} \right.,
   \label{eq:phil}
\end{equation}
Here we set Phillips' parameter $\alpha_\mathrm{P} = 0.0083$ (there
is some disagreement about its values with others workers, including
\citealt{holthuijsen07} and \citealt{webb11} preferring the value 0.0081). The
peak circular frequency is denoted $\omega_\mathrm{p}$.  
The Stokes drift profile under (\ref{eq:phil}) is
\begin{equation}
   v_\mathrm{P}(z) = 2 \int_{\omega_\mathrm{p}}^\infty \alpha_\mathrm{P} g
       \omega^{-2} \er^{2\omega^2z/g} \,\dr\omega.
   \label{eq:uphil}
\end{equation}
which can be found analytically [see eg \citealt{gradshteyn07}, Eq~(3.461.5)], 
\begin{equation}
   v_\mathrm{P}(z) = 2\alpha_\mathrm{P}g
   \left[\frac{1}{\omega_\mathrm{p}}\exp{(2\omega_\mathrm{p}^2z/g)} -
    \sqrt{-2\pi z/g}\left(1 - \mathrm{erf}(\omega_\mathrm{p}\sqrt{-2z/g})\right)
     \right].
     \label{eq:uphilsolved}
\end{equation}
The transport can also be found analytically,
\begin{equation}
   V_\mathrm{P} = \frac{\alpha_\mathrm{P}g^2}{3\omega_\mathrm{p}^3}.
   \label{eq:Uphil}
\end{equation}
Near the surface ($|z|$ small), the term involving the error function becomes
vanishingly small compared with the first term, and it is clear that
\begin{equation}
   v_\mathrm{P}(z) \approx \frac{2\alpha_\mathrm{P}g}{\omega_\mathrm{p}}
   \er^{2k_\mathrm{p}z}.
   \label{eq:uphilshallow}
\end{equation}
Here we have introduced the peak wavenumber $k_\mathrm{p} =
\omega_\mathrm{p}^2/g$.
To investigate the behavior for large $|z|$ we substitute the following
asymptotic expansion for the error function in \Eq{uphilsolved} [see
\cite{abr72}, Eq~(7.1.23)], valid for large $x$ (thus large $|z|$),
\begin{equation}
    \mathrm{erf}(x) \approx 1 - \frac{\er^{-x^2}}{x \sqrt{\pi}}\left(1 -
                       \frac{1}{2x^2}\right).
   \label{eq:erfasym}
\end{equation}
Hence, for large $|z|$ profile (\ref{eq:uphil}) drops off as
\begin{equation}
   v_\mathrm{P}(z) \approx
   \alpha_\mathrm{P}g^2 
   \frac{\er^{2k_\mathrm{p}z}}{ 2\omega_\mathrm{p}^3 |z|}.
   \label{eq:uphildeep}
\end{equation}
Motivated by this we have explored a profile which approaches
the exponential shape (\ref{eq:uphilshallow}) near the surface and 
which goes like the asymptotic solution (\ref{eq:uphildeep}) in the deep,
\begin{equation}
   v_\mathrm{e} = v_0 
    \frac{\er^{2k_\mathrm{\er}z}}{1-Ck_\mathrm{\er}z}.
    \label{eq:uve1}
\end{equation}
The coefficient that was found to minimize the MSE for the Phillips
spectrum over the entire profile is $C \approx 8$. Obviously the MSE takes
into account discrepancies over the entire water column and will be more
sensitive to deviations in the upper part where the drift is stronger.
The transport under such a profile involves the exponential integral $E_1$
and can be solved analytically [\citealt{abr72}, Eq~(5.1.28)] to yield
\begin{equation}
   V_\mathrm{s} = \frac{v_0 \er^{1/4} E_1(1/4)}{8 k_\mathrm{e}}.
   \label{eq:UVe} 
\end{equation}
It will in the following be referred to as the exponential integral profile.
This imposes the following constraint on the inverse depth scale,
\begin{equation}
   k_\mathrm{e} = \frac{v_0 \er^{1/4} E_1(1/4)}{8V_\mathrm{s}}.
   \label{eq:ke} 
\end{equation}
Here $\er^{1/4}E_1(1/4) \approx 1.34$, thus
\begin{equation}
   k_\mathrm{e} \approx \frac{{v}_0}{5.97V_\mathrm{s}} \approx
   k_\mathrm{m}/3. 
   \label{eq:keapprox} 
\end{equation}

\section{Profiles under Parametric Spectra}
\label{sec:paramspectra}
In the previous section we showed that the profile (\ref{eq:uve1}) approaches
the profile under the Phillips spectrum (\ref{eq:uphil}) near the surface
(\ref{eq:uphilshallow}) and in the deep (\ref{eq:uphildeep}).  We will
now assess the quantitative and qualitative differences between the two
approximate profiles, referred to by subscripts ``m'' for monochromatic and
``e'' for exponential integral, with respect to parametric spectra.

\begin{figure}[h]
\begin{center}
(a)\includegraphics[scale=0.6]{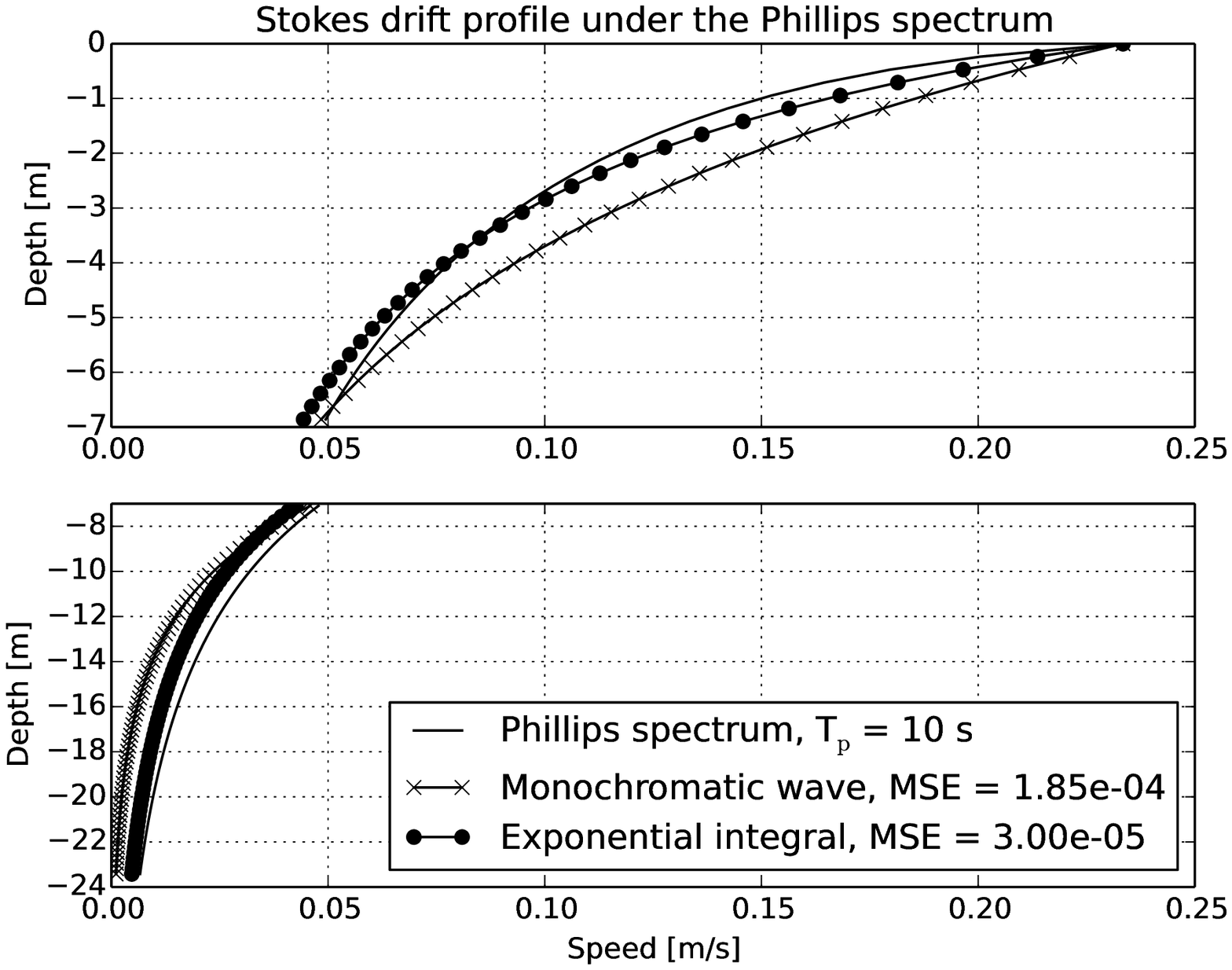}\\
(b)\includegraphics[scale=0.6]{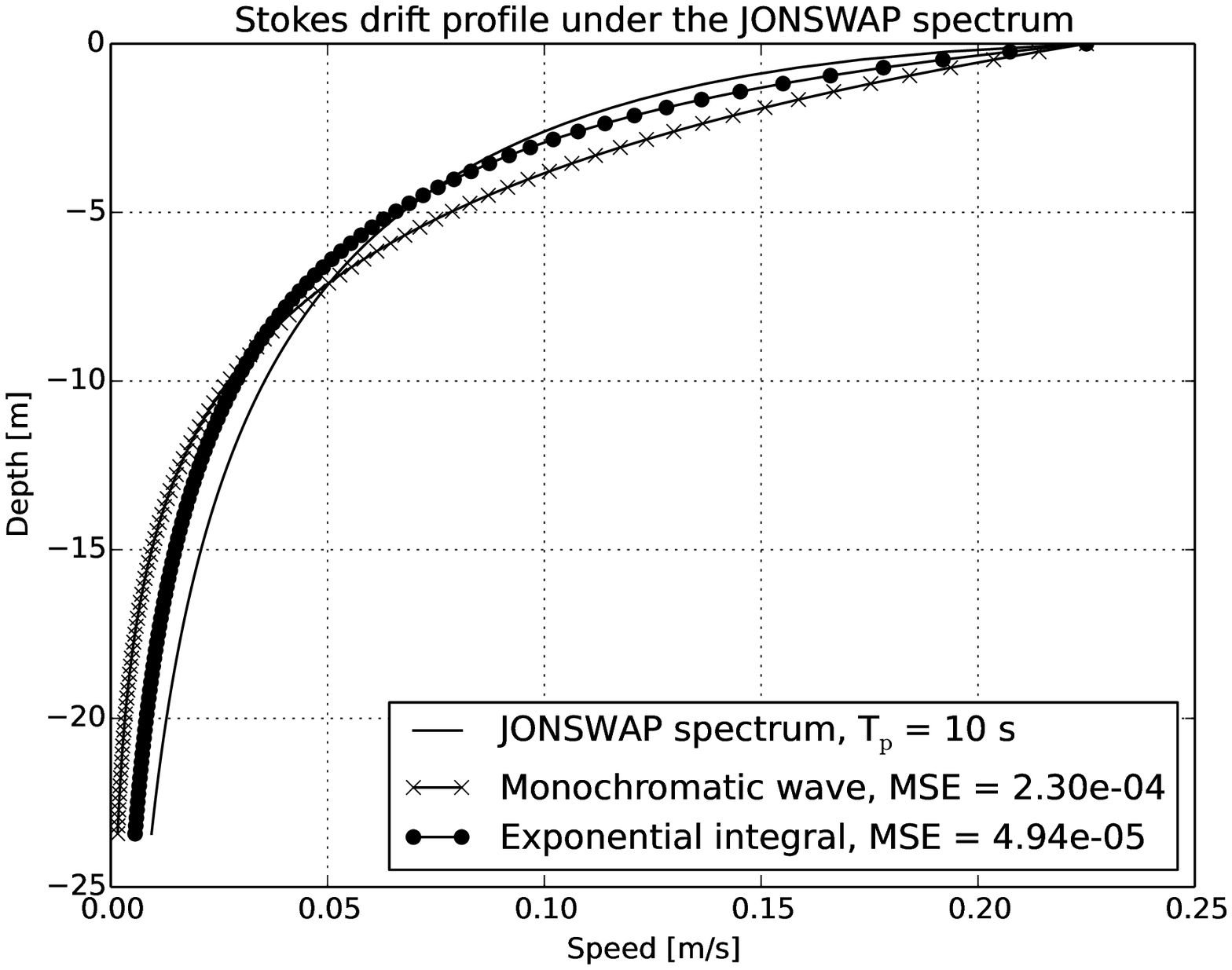}\\
\caption{Panel a: The Stokes drift profile under the Phillips spectrum
($T_\mathrm{p} = 10\, \mathrm{s}$). The upper part of Panel a is a zoom
of the upper 7 meters for readability.  The monochromatic approximation
(x) tends to overestimate the drift in the upper part of the water column
while underestimating the drift in the deeper part. The exponential integral
approximation (o) exhibits closer correspondence throughout the water column,
with an MSE about six times smaller than that found for the monochromatic
approximation.  
Panel b: The Stokes drift profile under the JONSWAP spectrum
($T_\mathrm{p} = 10\,\mathrm{s}$, fetch $X = 10\,\mathrm{km}$). The results
are similar to those for the Phillips spectrum with an MSE of the
exponential integral (o) about 60\% smaller than that of the monochromatic
approximation (x).}
\label{fig:param_profiles}
\end{center}
\end{figure}

The profile under the Phillips spectrum (\ref{eq:uphil}) is compared with
the two approximate profiles (\ref{eq:uvmono}) and (\ref{eq:uve1}) in Panel
a of \Fig{param_profiles}. The exponential integral approximation has an
MSE of about a sixth that of the monochromatic approximation. As
mentioned in the previous section, the coefficient used for the fit is found
by minimizing the MSE with respect to the Phillips spectrum. It
is therefore not surprising that the match is good for this spectrum.

The Pierson-Moskowitz (P-M) spectrum \citep{pie64} is commonly used to model
fully developed (equilibrium) sea states,
\begin{equation}
   F_\mathrm{PM} = 
     \alpha_\mathrm{P} g^2 \omega^{-5} \exp
     \left[-\frac{5}{4}\left(\frac{\omega_\mathrm{p}}{\omega}\right)^4\right].
     \label{eq:PM}
\end{equation}
We find the same general improvement as was found for the Phillips spectrum
above with an MSE about a fifth that of the monochromatic approximation
(not shown).  Note that here the integral covers also the lower frequencies as
the spectrum remains bounded for all frequencies.  Panel b shows the profile
under the JONSWAP spectrum.  This spectrum is based on the P-M spectrum with
a peak enhancement to account for the spectral shape found in fetch-limited
seas \citep{has73,jan04,webb11}
\begin{equation}
   F_\mathrm{JONSWAP} = F_\mathrm{PM} \gamma^\Gamma,
     \label{eq:JONSWAP}
\end{equation}
where
\begin{equation}
   \Gamma =
   \exp \left[-\frac{1}{2}\left(\frac{f/f_\mathrm{p}-1}{\sigma}\right)^2\right]
     \label{eq:Gamma}.
\end{equation}
Here typical values are $\gamma = 3.3$, $\sigma = 0.07$ for $f \leq
f_\mathrm{p}$ and $\sigma = 0.09$ when $f > f_\mathrm{p}$.  The exponential
integral profile gives a reduction in MSE of about 60\% compared with the
monochromatic profile (see \Fig{param_profiles}b).

\subsection{The Shear of the Stokes Drift Profile}
\label{sec:shear}
The production of Langmuir turbulence arises from a vortex force term,
$\mathbf{v}_\mathrm{s} \times \nabla \times \mathbf{u}$, in the momentum
equation \citep{leibovich83}. It is assumed that the vortex force gives rise
to a term involving the shear of the Stokes drift velocity profile
in the turbulence kinetic energy [\citet{sky95}, \citet{mcwilliams97},
\citet{teixeira02}, \citet{kantha04}, \citet{ardhuin06}, \citet{polton07},
\citet{grant09} and \citet{belcher12}], although it is somewhat unclear
whether this effect will be strong enough to explain the observed Langmuir
circulation.  The turbulent kinetic energy (TKE) equation with a Stokes drift
shear term can be written
\begin{equation}
     \frac{\Dr e}{\Dr t} =
      \frac{g}{\rho_\mathrm{w}}\overline{u'_3 \rho'} 
     -\UR \Dp{\overline{u}_i}{x_j} 
     -\UR \Dp{v_i}{x_j} 
     -\Dp{}{x_j}(\overline{u'_je})
     -\frac{1}{\rho_\mathrm{w}}\Dp{}{x_i}(\overline{u'_ip'})
     -\epsilon.
     \label{eq:tker}
\end{equation}
Here, $e \equiv q^2/2 = \overline{u'_i u'_i}/2$ is the TKE per unit
mass (with $q$ the turbulent velocity) and $\epsilon$ is the dissipation
[see e.g. \citet{stu88} p 152].  The term involving the Reynolds stresses
multiplied by the gradient in Stokes drift velocity, $v_i$, represents
production of Langmuir turbulence \citep{mcwilliams97,teixeira02,ardhuin06}.
By making the gradient transport closure approximation \citep{stu88,janssen12},
ignoring advective terms and horizontal gradients, and rewriting in vectorial
form we arrive at
\begin{equation}
   \Dp{e}{t} = 
   -\nu_\mathrm{h} N^2 
   +\nu_\mathrm{m} S^2 
   +\nu_\mathrm{m} \mathbf{S} \cdot \Dp{\mathbf{v}_\mathrm{s}}{z} 
   -\Dp{}{z}(\overline{w'e}) 
   -\frac{1}{\rho_\mathrm{w}}\Dp{}{z}(\overline{w'p'})
   -\epsilon.
     \label{eq:tke}
\end{equation}
Here we have reverted to using $z$ for the vertical axis and $w$ for
vertical velocities.  We recognize in Eqs~(\ref{eq:tker})--(\ref{eq:tke})
the familiar terms of the TKE equation [see \citealt{stu88}, Eq (5.1a)],
namely shear production, $S^2 = (\partial \overline{\mathbf{u}}/\partial
z)^2$, and buoyancy production through the Brunt-Vais\"{a}l\"{a} frequency,
$N^2 = -(g/\rho) \dr\rho/\dr z$ ($\nu_\mathrm{m,h}$ are turbulent 
viscosity and diffusion coefficients, respectively) as well as the divergences
of the pressure correlation term $\overline{w'p'}$ and turbulent transport
$\overline{w'e}$.

It is of interest to investigate the shear under parametric spectra, and for
the Phillips spectrum (\ref{eq:uphil}) an analytical solution can be found
[\citealt{gradshteyn07}, Eq~(3.321.2)],
\begin{equation}
   \Dp{v_\mathrm{P}}{z} = 
     2 \alpha_\mathrm{P} g \int_{\omega_\mathrm{p}}^\infty
       \er^{-2\omega^2|z|/g} \,\dr\omega 
     = \sqrt{\frac{\pi g}{8|z|}} \mathrm{erfc}\left(\sqrt{\frac{2|z|}{g}
     \omega_\mathrm{p}}\right).
   \label{eq:philshear}
\end{equation}
On the surface the shear goes to infinity.
This is in contrast to the shear under a monochromatic wave (\ref{eq:uvmono}),
which remains bounded near the surface,
\begin{equation}
   \Dp{v_\mathrm{m}(z=0)}{z} = 2k_\mathrm{m} v_0.
   \label{eq:monoshear}
\end{equation}
The shear of the exponential integral profile (\ref{eq:uve1}) also remains
bounded, but reaches a value approximately 67\% higher than the monochromatic
profile at the surface,
\begin{equation}
   \Dp{v_\mathrm{e}(z=0)}{z} = 10k_\mathrm{e} v_0 \approx 
     \frac{10}{3}k_\mathrm{m} v_0.
   \label{eq:expshear}
\end{equation}
Technically the singularity in \Eq{philshear} can be avoided by moving the
computation of the Stokes shear away from the surface through the use of a
staggered grid. It is also evident that for real ocean waves the spectrum
will not extend to infinite wavenumbers \citep{elfouhaily97}. In practice,
though, it may be necessary to cap the Stokes shear near the surface when
estimating the Langmuir turbulence when assuming a tail proportional to
$f^{-5}$ (see next Section).

\section{High-frequency contribution to the profile}
\label{sec:wamspectra}
The same procedure as outlined in \Sec{paramspectra} can be used to
compute the profiles and transports from discretized wave spectra with a
high-frequency cut-off. However, as the Stokes drift is weighted toward the
high-frequency (HF) part of the spectrum, the tail beyond the cut-off frequency
($f_\mathrm{c}$) is significant both for the profile and the transport. We
follow \citet{kom94} pp 233--234 and assume a tail of the form
\begin{equation}
   F_\mathrm{HF} =
   F(f_\mathrm{c},\theta)\left(\frac{f_\mathrm{c}}{f}\right)^5,
   \label{eq:Fhf}
\end{equation}
which is consistent with the Phillips spectrum (\ref{eq:phil}).  The
two-dimensional spectrum below the cut-off frequency is here assumed to
come from observations or from a numerical wave prediction model.  This is
the procedure used for adding the diagnostic high-frequency contribution
to the spectrum in the WAM model, see \citealt{has88,kom94,jan04}
as well as the WaveWatch-III model, \citealt{tol91,tolman02}).  In the ECMWF
version of the WAM model (ECWAM, see \citealt{wam40r1} for further details),
a lower diagnostic cut-off is set at
\begin{equation}
   f_\mathrm{d} = \min(f_\mathrm{max}, 2.5\overline{f}_\mathrm{windsea}).
   \label{eq:fd}
\end{equation}
Here, $\overline{f}_\mathrm{windsea}$ is the mean frequency of windsea based on
the first moment and $f_\mathrm{max}$ is the highest resolved frequency of the
modeled spectrum. Above $f_\mathrm{d}$ the spectrum is treated diagnostically,
ie, a tail of the form (\ref{eq:Fhf}) overwrites the prognostic tail.

The high-frequency tail adds the following contribution,
\begin{equation}
   \mathbf{v}_\mathrm{HF}(z) = \frac{16\pi^3}{g}f_\mathrm{c}^5
   \int_0^{2\pi} F(f_\mathrm{c},\theta) \hat{\mathbf{k}} \, \dr\theta 
   \int_{f_\mathrm{c}}^{\infty} \frac{\exp{(8\pi^2zf^2/g)}}{f^2} \,\dr f.
   \label{eq:utail}
\end{equation}
The latter integral is similar to (\ref{eq:uphil}) and can be solved in a similar manner to (\ref{eq:uphilsolved})  [see eg \citealt{gradshteyn07}, Eq~(3.461.5)], yielding
\begin{equation}
   \mathbf{v}_\mathrm{HF}(z) = \frac{16\pi^3}{g}f_\mathrm{c}^5 
    \int_0^{2\pi} F(f_\mathrm{c},\theta) \hat{\mathbf{k}} \, \dr\theta
    \left[\frac{\exp{(-\mu f_\mathrm{c}^2)}}{f_\mathrm{c}} - 
     \sqrt{\mu\pi}\left(1-\mathrm{erf}(f_\mathrm{c}\sqrt{\mu})\right)\right],
     \label{eq:uhf}
\end{equation}
where $\mu = -8\pi^2z/g$.  The high-frequency addition to the surface Stokes
drift in deep water is
\begin{equation}
   \mathbf{v}_\mathrm{HF}(0) = \frac{16\pi^3}{g}f_\mathrm{c}^4 
    \int_0^{2\pi} F(f_\mathrm{c},\theta) \hat{\mathbf{k}} \, \dr\theta.
     \label{eq:uhf0}
\end{equation}
ECWAM \citep{wam40r1} computes and outputs the surface Stokes drift velocity
vector corrected for the high-frequency contribution.  The tail contribution
to the transport is
\begin{equation}
   \mathbf{V}_\mathrm{HF} = \frac{2\pi}{3}f_\mathrm{c}^2
   \int_0^{2\pi} F(f_\mathrm{c},\theta) \hat{\mathbf{k}} \, \dr\theta.
   \label{eq:Uhf}
\end{equation}

\section{Modeled Profiles in the North Atlantic}
\label{sec:erai}
The ERA-Interim is a continuously updated atmospheric and wave field reanalysis
produced by ECMWF, starting in 1979.  The model and data assimilation scheme
of the reanalysis are based on Cycle 31r2 of the Integrated Forecast System
(IFS). ECWAM is coupled to the atmospheric part of the IFS
(see \citealt{jan04} for details of the coupling and \citealt{dee11}
for an overview of the ERA-Interim reanalysis).  The resolution of the
wave model component is $1.0^\circ$ on the Equator but the resolution is
kept approximately constant globally through the use of a quasi-regular
latitude-longitude grid where grid points are progressively removed toward the
poles \citep{jan04}. A similar scheme applies for the atmospheric component,
but here the resolution is approximately $0.75^\circ$ at the Equator. The
wave model is run with shallow water physics where appropriate.  The spectral
range from $3.45\times10^{-2}$ to $0.55 \, \mathrm{Hz}$ is spanned with 30
logarithmically spaced frequency bands. The angular resolution is $15^\circ$.

\begin{figure}[h]
\begin{center}
\hspace{5mm}
\includegraphics[scale=0.8]{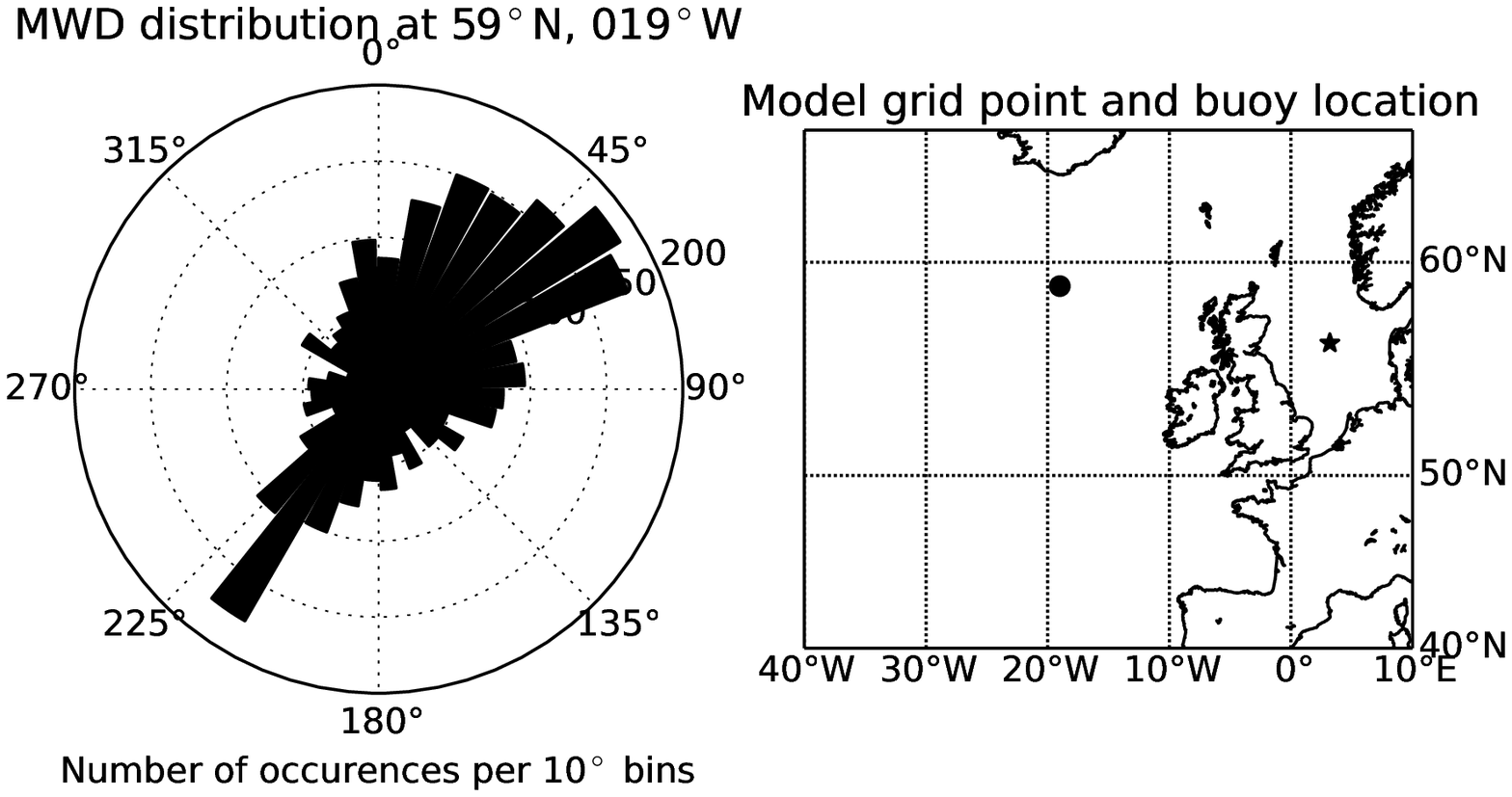}\\
\caption{Left panel: The directional distribution of the mean wave direction
(going to) in model location 59$^\circ$~N, 019$^\circ$~W. A large 
spread in wave direction is found. The location has a high prevalence of wind 
sea but is also exposed to swell. Right panel: Model location (circle) and buoy
location (*) at 56.5$^\circ$N, 003.2$^\circ$E.}
\label{fig:mwd_hist_and_location}
\end{center}
\end{figure}

\begin{figure}[h]
\vspace{4mm}
\begin{center}
\hspace{5mm}
\includegraphics[scale=0.9]{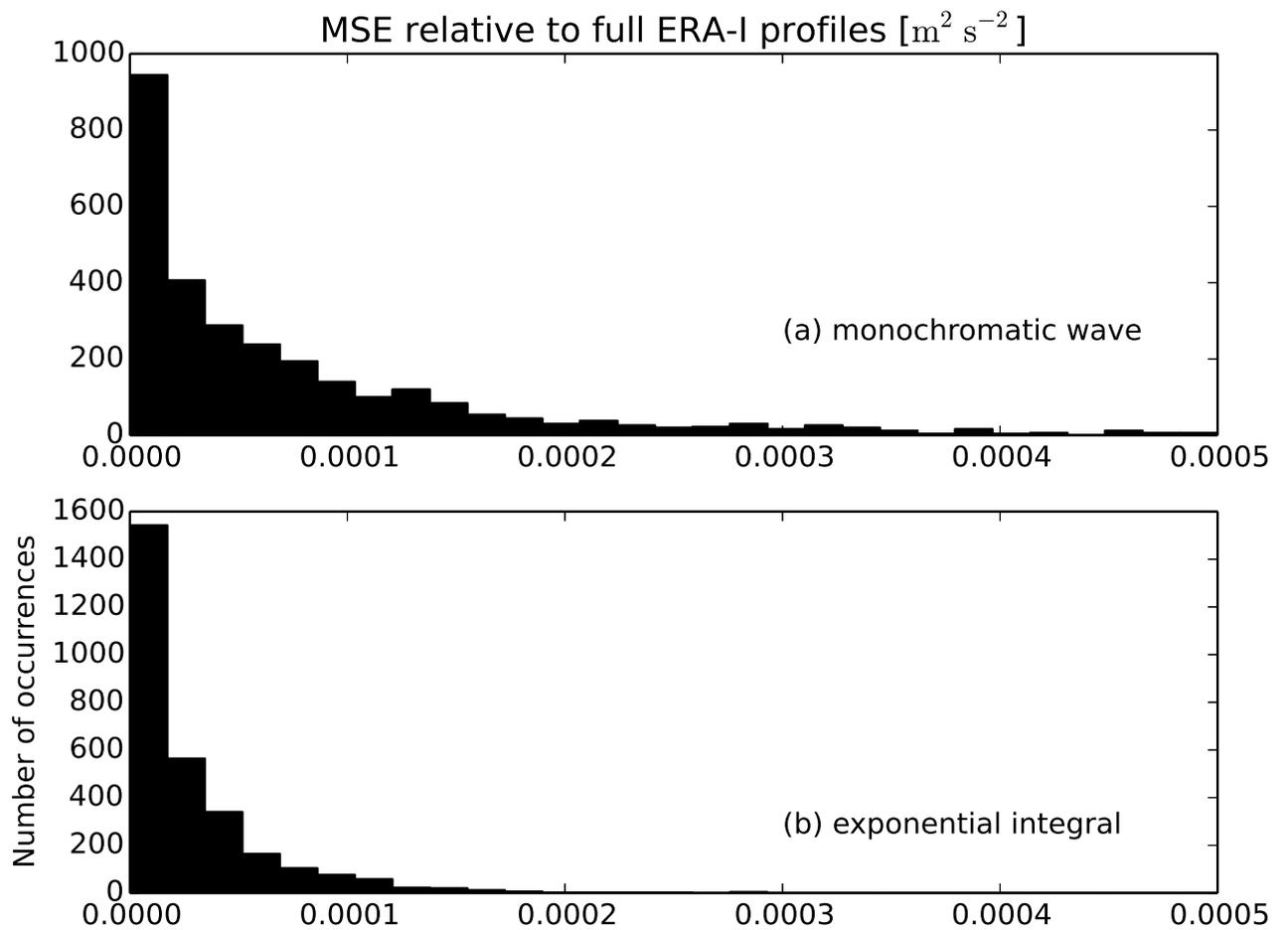}\\
\vspace{4mm}
\caption{Panel a: The MSE between the full Stokes
profile and the monochromatic profile to 30 m depth (vertical resolution
0.1 m).  Panel b: The MSE of the exponential integral profile is
on average about one third that of the monochromatic profile shown in Panel a.}
\label{fig:stats_hist}
\end{center}
\end{figure}

For this study we computed the Stokes drift profiles down to 30~m depth from
the two-dimensional ERA-Interim spectra in a region in the north Atlantic
ocean ($59-60^\circ$N, $20-19^\circ$W, see \Fig{mwd_hist_and_location})
for the whole of 2010. This region is stormy while also exposed to swell,
providing a range of complex wave spectra.  To assess the difference between
the monochromatic approximation and the exponential integral approximation
the MSE from the full Stokes drift profile to 30~m depth was
calculated for every spectrum. The results are shown in \Fig{stats_hist}.

\begin{figure}[h]
\begin{center}
\begin{tabular}{cc}
 \includegraphics[scale=0.35]{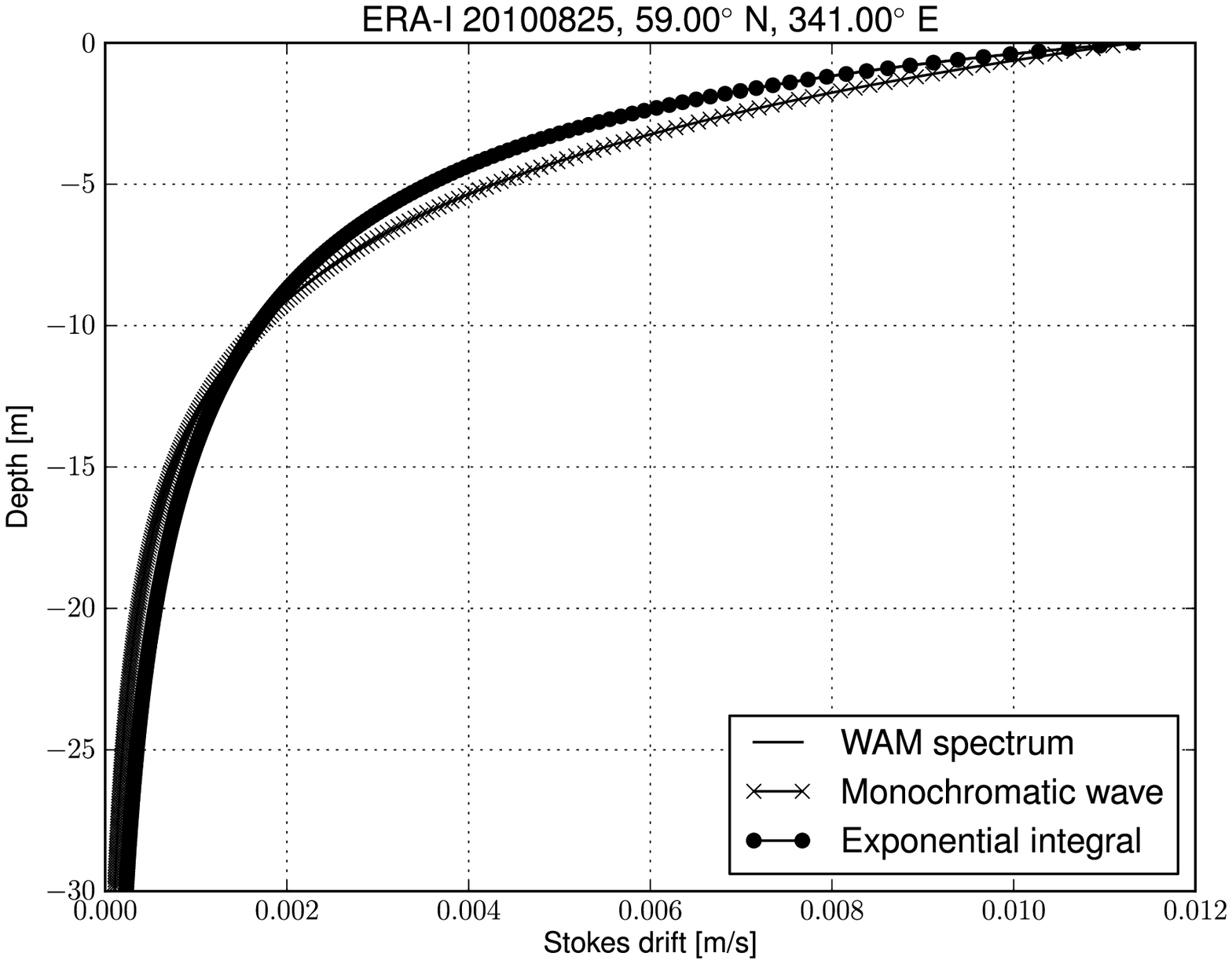}&
 \includegraphics[scale=0.35]{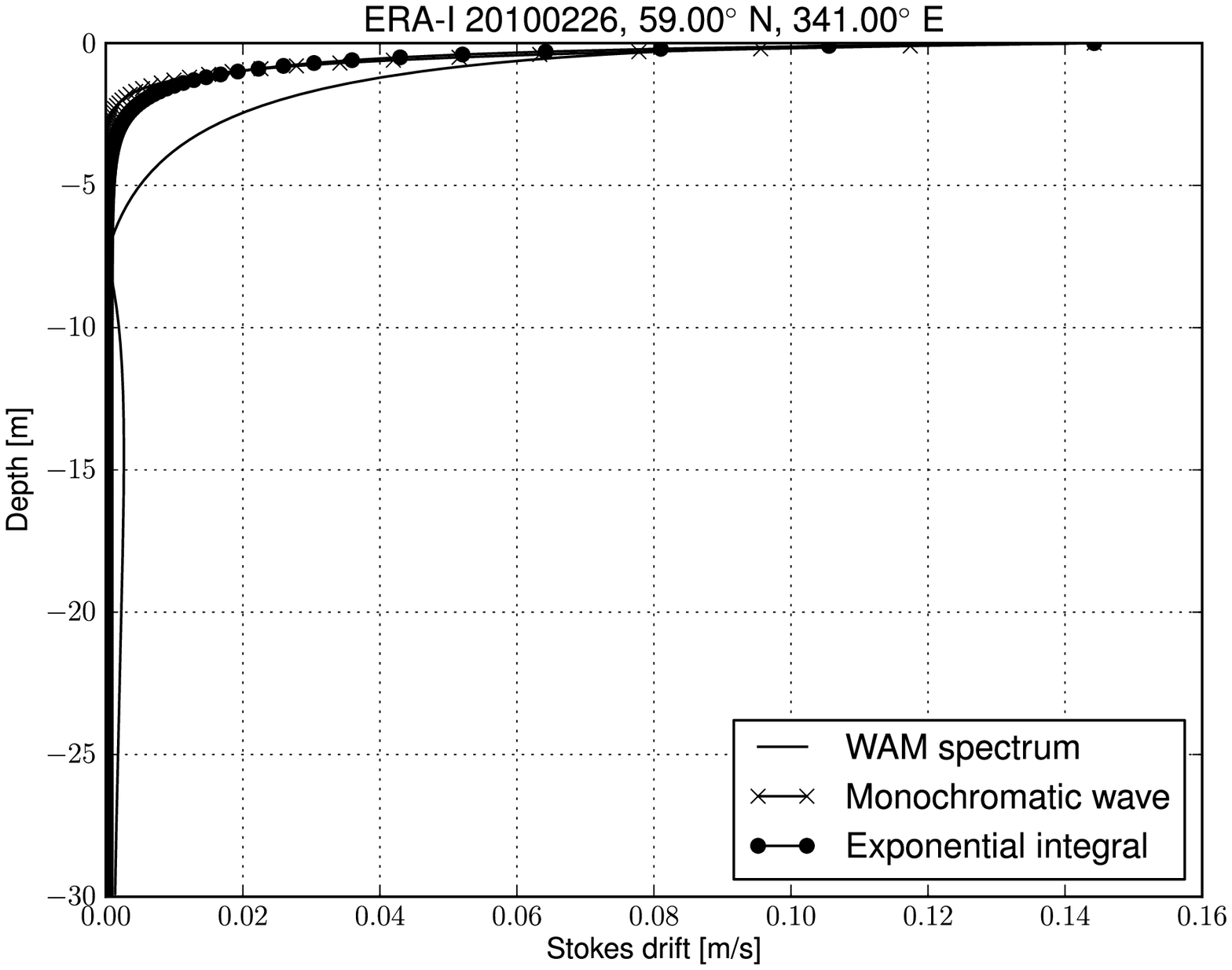}\\
(a)\includegraphics[scale=0.35]{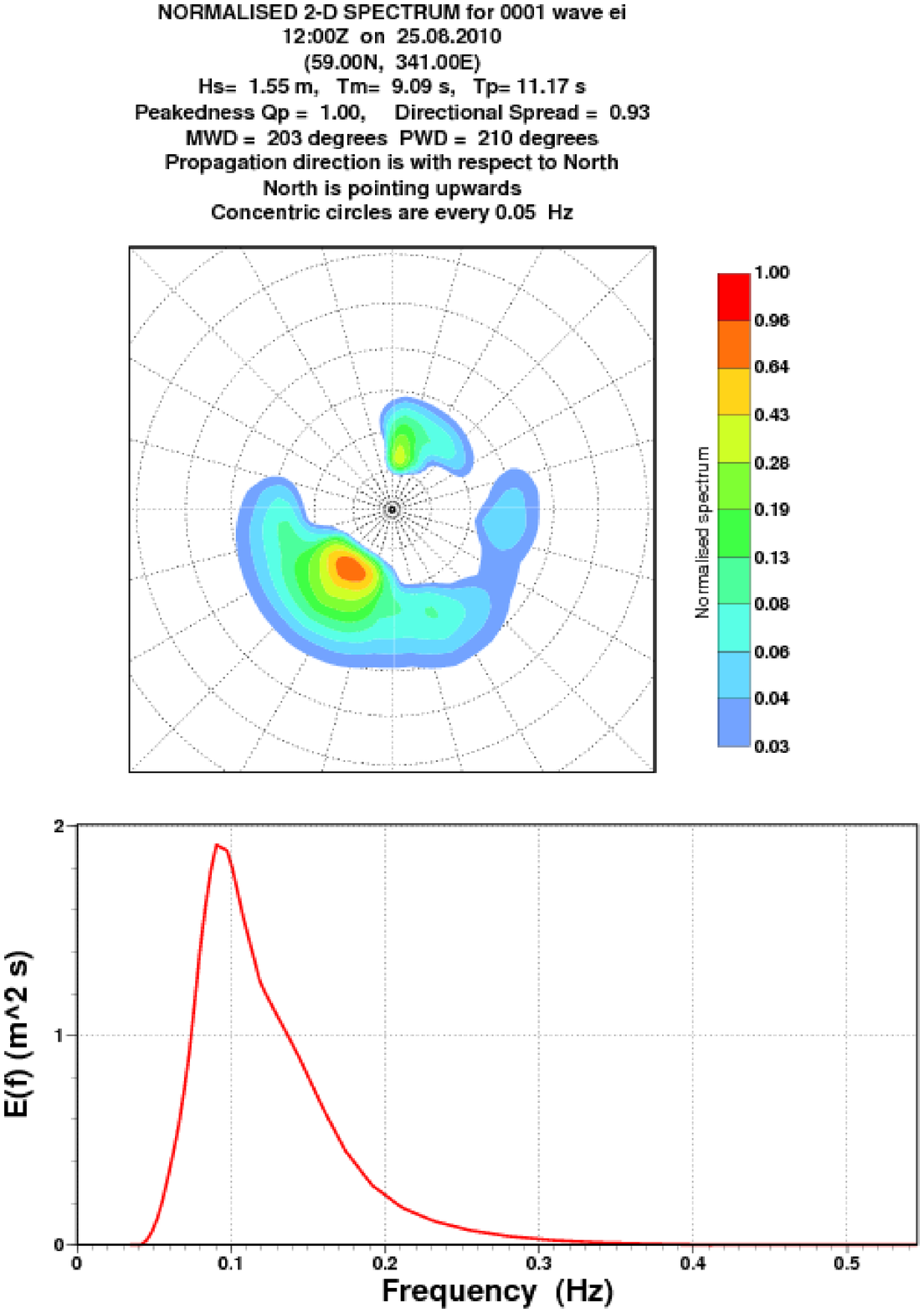}&
(b)\includegraphics[scale=0.35]{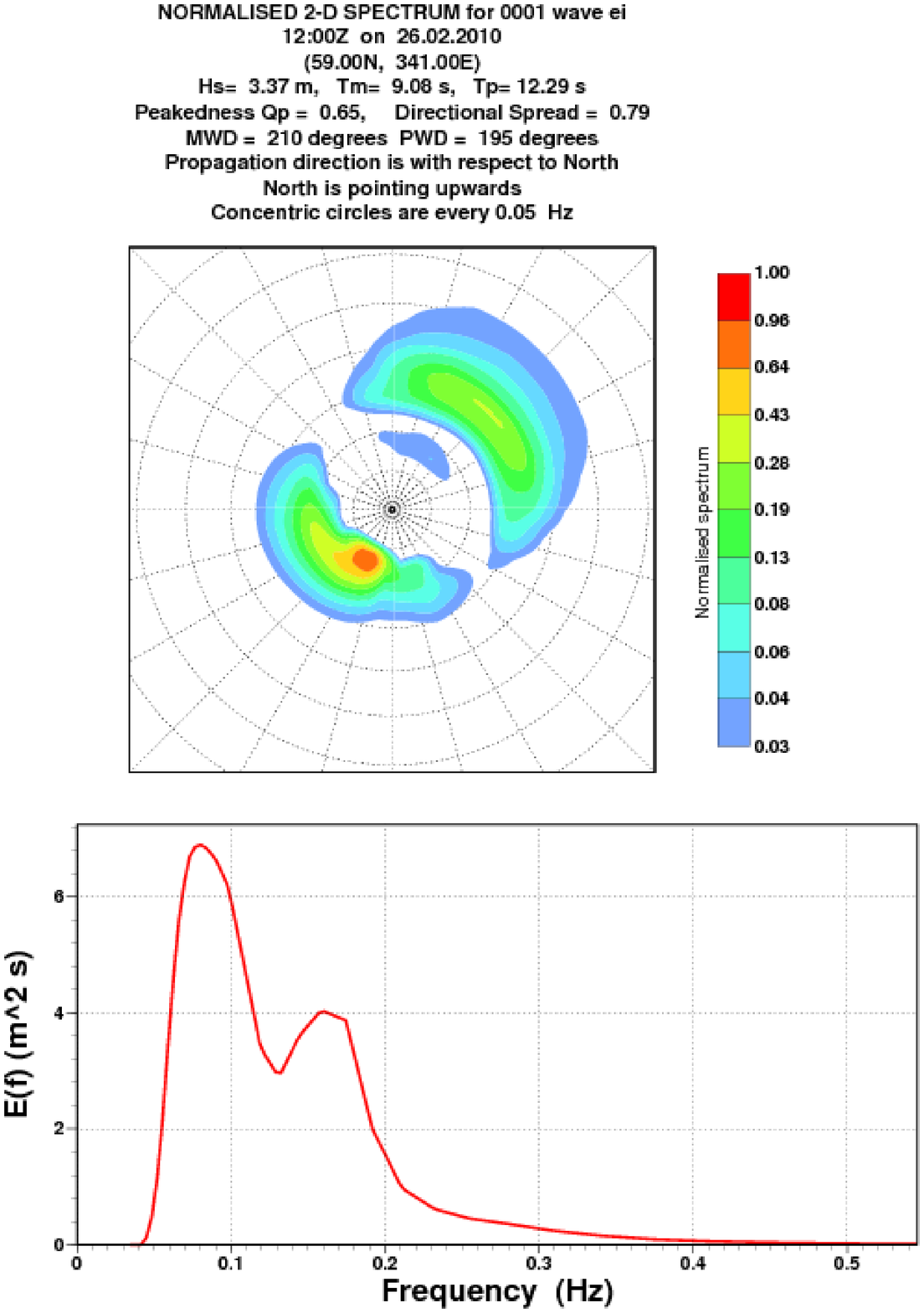}\\
\end{tabular}
\caption{Panel a: The Stokes drift profile under a full two-dimensional wave spectrum
from the ERA-Interim reanalysis. The location is in the north Atlantic.
An extremely good fit is found in this case. The 2-D spectrum shows a strong
bimodality which is masked in the 1-D spectrum.
Panel b: Much poorer fit is found in this case where a strong swell system is
superimposed on locally generated wind sea. There is still some improvement over
the monochromatic approximation. Here the swell part is dominant and of a lower
frequency, making the 1-D spectrum bimodal.}
\label{fig:erai_profile}
\end{center}
\end{figure}

The MSE of the exponential integral profile from the full Stokes
profile is on average 35\% that of the monochromatic profile for our
chosen location and model period (2010).  The improvement is consistent
for a range of different sea states, as illustrated in \Fig{erai_profile}.
In Panel a the match is so close that the exponential integral profile
overlaps the full profile.  Poor performance is expected in cases where a
one-dimensional fit is made to wave spectra with two diametrically opposite
wave systems. Such a case is shown in Panel b, where a swell system travels
in the opposite direction of the wind sea.  Indeed, this spectrum represents the
worst fit found throughout the model period, but even here there is slight
improvement over the monochromatic approximation.

\subsection{Tail-sensitivity of Modeled Stokes Drift Profiles}
\begin{figure}[h]
\begin{center}
(a)\includegraphics[scale=0.6]{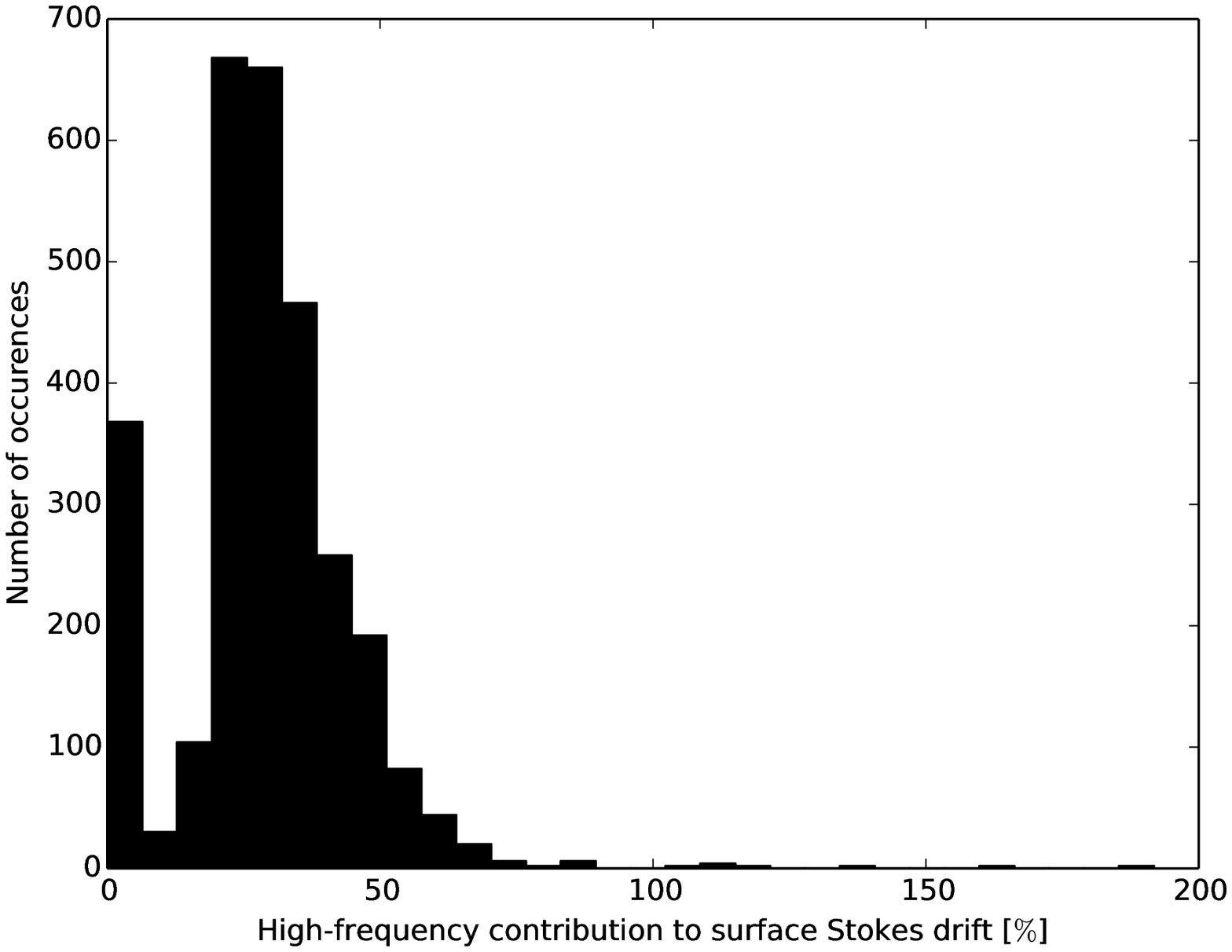}\\
(b)\includegraphics[scale=0.6]{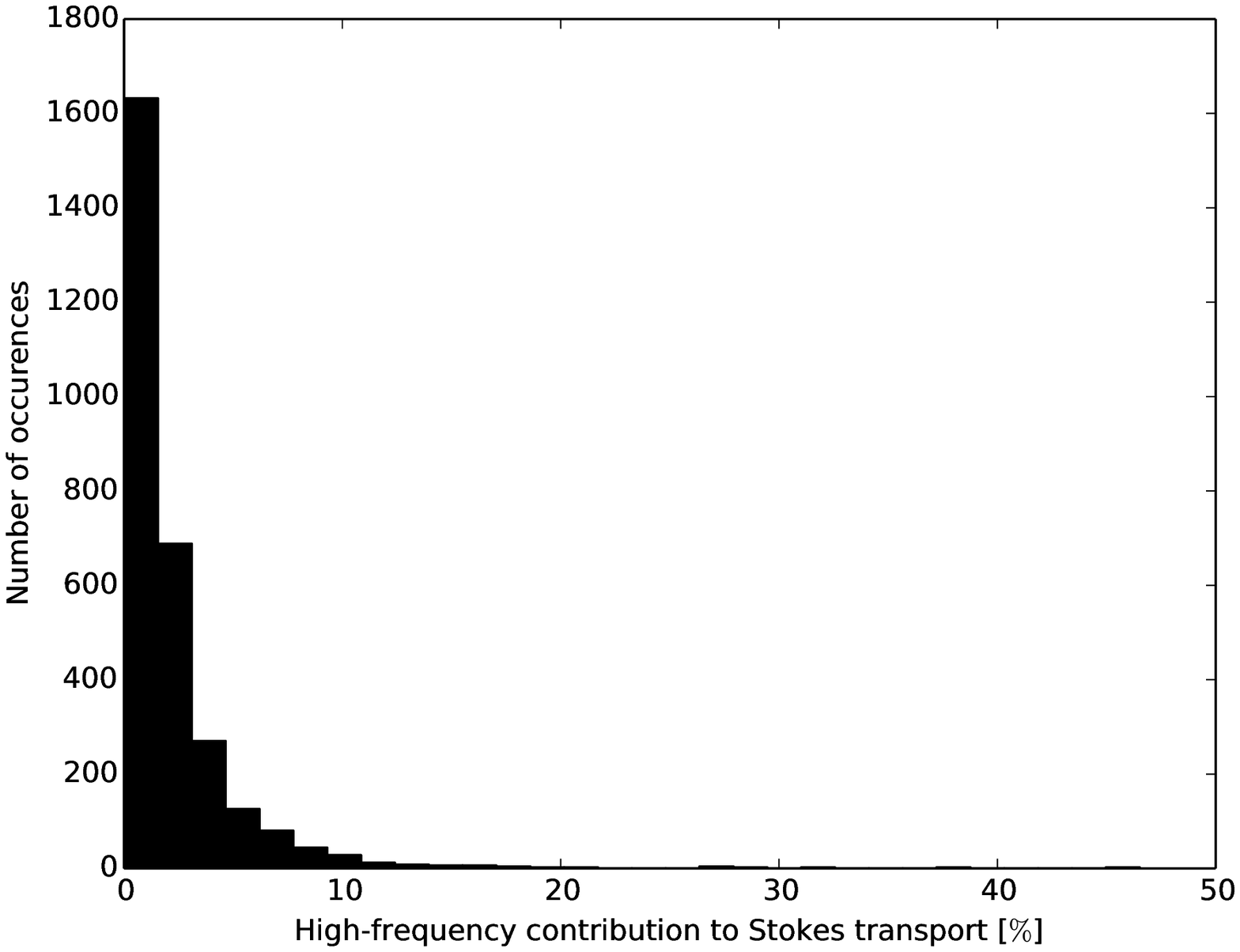}\\
\caption{
Panel a: Ratio of high-frequency contribution to the surface Stokes drift. On
average the contribution is about 39\%.  
Panel b: Ratio of high-frequency
contribution to the Stokes transport. On average the contribution is about
3\%, and only occasionally will it exceed 10\%.}
\label{fig:uhf}
\end{center}
\end{figure}

It is well known that adding the contribution from the high-frequency tail is
important, and indeed it is standard practice to include it in the computation
of the surface Stokes drift velocity (see eg the ECMWF model documentation,
\citealt{wam40r1}, p 52).  We find that the contribution from the spectral
tail to the surface Stokes drift velocity found in \Eq{uhf} on average is
about a third, and sometimes exceeding 75\% (\Fig{uhf}, Panel a).
In contrast, its contribution to the \emph{transport} (\ref{eq:Uhf}) is
generally small (average 3\%, Panel b, \Fig{uhf}), although in certain
cases it may exceed 10\%.

\begin{figure}[h]
\begin{center}
(a)\includegraphics[scale=0.6]{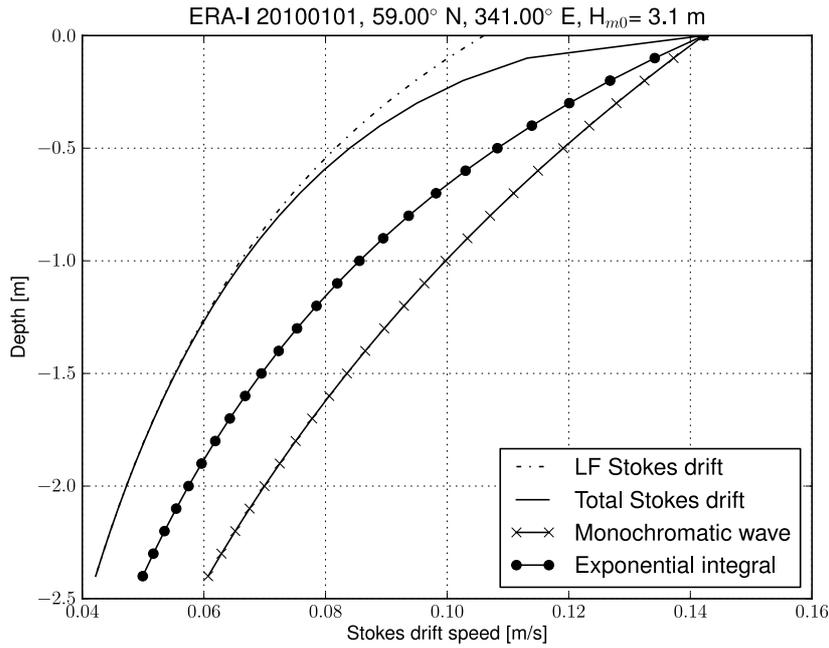}\\
(b)\includegraphics[scale=0.6]{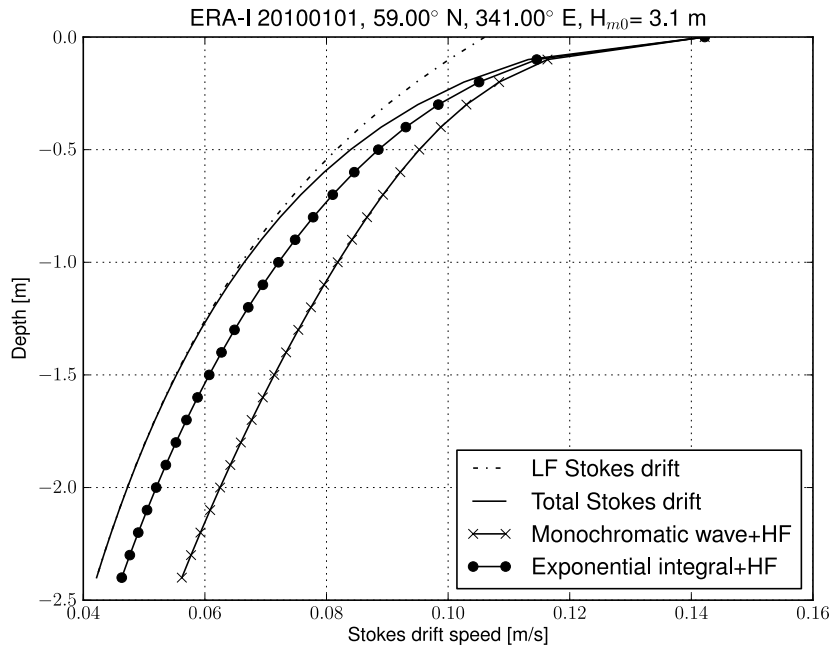}\\
\caption{Panel a: The high-frequency contribution to the Stokes drift velocity. The
Short waves beyond the cut-off frequency contribute only to the drift in the
upper half meter (compare the dash-dotted low-frequency Stokes drift to the
total drift drawn with a full line). The two
approximate profiles are pegged to the surface Stokes drift and coincide
exactly at the surface. The shear is not well represented by either of the
approximate profiles in the upper half meter, but the exponential integral
profile is the better match of the two.
Panel b: The same approximate profiles with the high-frequency profile added. A
much better match for the upper meters of the ocean is achieved, both in terms
of shear and absolute error.}
\label{fig:uhf_profile}
\end{center}
\end{figure}

The high-frequency contribution decays rapidly with depth, as can be seen
in Panel a of \Fig{uhf_profile}.  Below 0.5 m the difference between the
low-frequency (LF) profile and the full profile is negligible. Neither of
the approximate profiles is a particularly good match, but of the two
the exponential integral profile has a slightly better gradient than
the monochromatic profile. This mismatch in the upper half meter is in
contrast to the good overall match found for the whole water column (see
\Fig{erai_profile}). Panel b shows the approximate profiles instead pegged to the low-frequency surface Stokes drift with the high-frequency contribution added after. Now the gradient is much closer to that of the theoretical full Stokes profile, with the exponential integral profile being a good match. 
In principle it is straightforward to add this contribution to the approximate
profile by way of \Eq{uhf}, but it requires knowledge of the two-dimensional
wave spectrum at the cut-off frequency $f_\mathrm{c}$.

These results are to some extent dependent on the way the tail is
formulated [see Eqs~(\ref{eq:Fhf})-(\ref{eq:fd}) and \citet{wam40r1}, p 52].
\citet{ardhuin09} argues that the high-frequency fit to buoy data is better
when using a revised dissipation source function rather than the formulation
presented by \citet{bidlot07} which is currently employed by ECWAM. However,
the analytical form of the approximate profile has been shown to fit analytical
and empirical spectra well (cf \Sec{paramspectra}), and it seems unlikely
that a revised dissipation will seriously change the results found here.

\subsection{Discrepancy Between the Stokes Transport and $m_1$}
It is clear that 
\begin{equation}
   |\mathbf{V}_\mathrm{s}| \leq 2\pi m_1,
\end{equation}
but it is not clear how large this deviation is on average for 
typical wave spectra in the open ocean. Assessing the overestimation is of 
practical value since the first spectral moment is often archived or indirectly
measured. Since the mean frequency is defined as $\overline{f} = m_1/m_0$
\citep{wmo98,holthuijsen07} and the significant wave height $H_{m_0} =
4\sqrt{m_0}$, we can derive the first moment from
the integrated parameters of a wave model or from wave observations and find
an estimate for the Stokes transport,
\begin{equation}
  \mathbf{V}_\mathrm{s} \approx \frac{2\pi}{16} \overline{f} H_{m_0}^2
  \hat{\mathbf{k}}_\mathrm{s}.
  \label{eq:UVHsf}
\end{equation}
Here $\hat{\mathbf{k}}_\mathrm{s} = (\sin \theta_\mathrm{s}, \cos
\theta_\mathrm{s})$ is the unit vector in the direction $\theta_\mathrm{s}$
of the Stokes transport. 

\begin{figure}[h]
\begin{center}
\hspace{5mm}
(a)\includegraphics[scale=0.6]{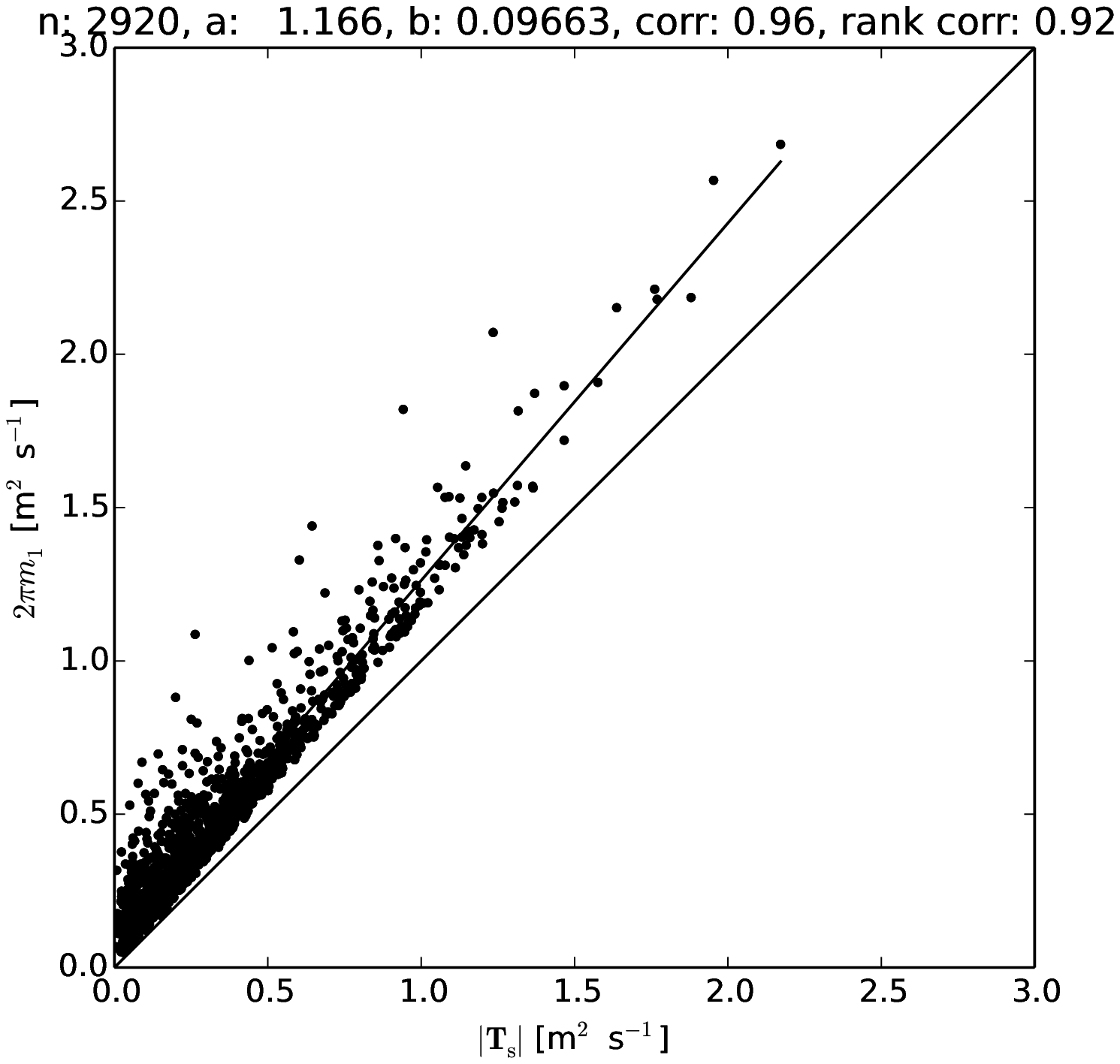}\\
(b)\includegraphics[scale=0.6]{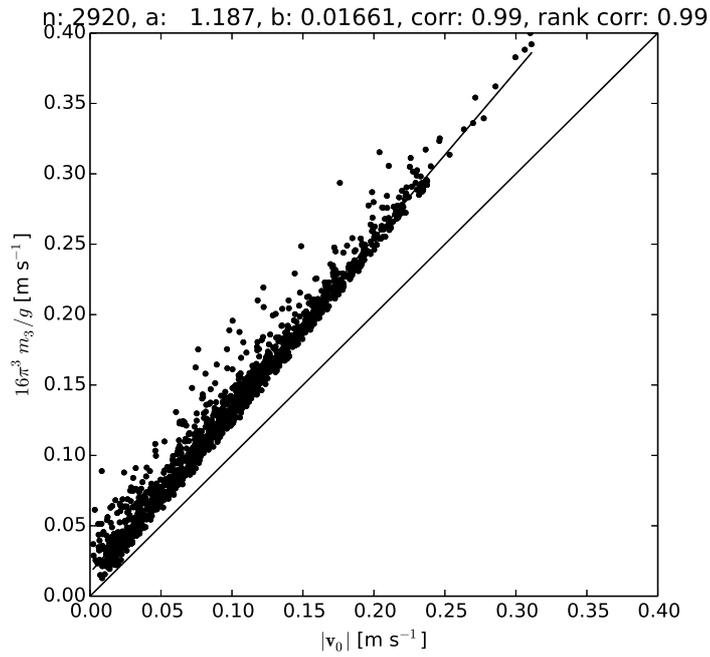}\\
\caption{Panel a: The discrepancy between the two-dimensional Stokes transport 
$|\mathbf{V}_\mathrm{s}|$ and the unidirectional estimate $2\pi m_1$
from the ERA-Interim reanalysis. Good agreement is generally found, but the
unidirectional estimate will on average be 16\% too high.
Panel b: The discrepancy between the two-dimensional surface Stokes drift
$|\mathbf{v}_0|$ and the unidirectional estimate $16\pi^3m_3/g$. The
unidirectional estimate will on average be 18\% too high.}
\label{fig:stokes_v_moments}
\end{center}
\end{figure}

Note that this Stokes transport direction is not normally archived by wave
prediction models (an exception being the more than 20-year long hindcast data set presented by \citealt{rascle13}), but it can be approximated by the mean wave direction
$\overline{\theta}$ as will be shown later.  Estimating the Stokes transport
from the first moment is attractive since it involves only integrated
parameters readily available from wave models.  \Fig{stokes_v_moments}a shows good
correspondence between the the Stokes transport and the estimate based on
$m_1$ in \Eq{UVHsf} with a correlation coefficient of $0.96$, but $m_1$ will
overestimate the transport on average by $17\%$.  Both transport estimates
include the contribution from the diagnostic high-frequency spectral tail.
Similarly, the surface Stokes drift velocity 
\begin{equation}
  |\mathbf{v}_0| \leq 16\pi^3m_3/g.
  \label{eq:stokes_m3}
\end{equation}
The estimate from $m_3$ will on average be about $19\%$ too high
(\Fig{stokes_v_moments}b). This is very close to the number reported by
\citet{ardhuin09} in their Appendix C (a reduction factor of 0.84).

\subsection{Deviation between the Stokes transport direction and the mean wave
direction} 
\begin{figure}[h]
\begin{center}
(a)\includegraphics[scale=0.6]{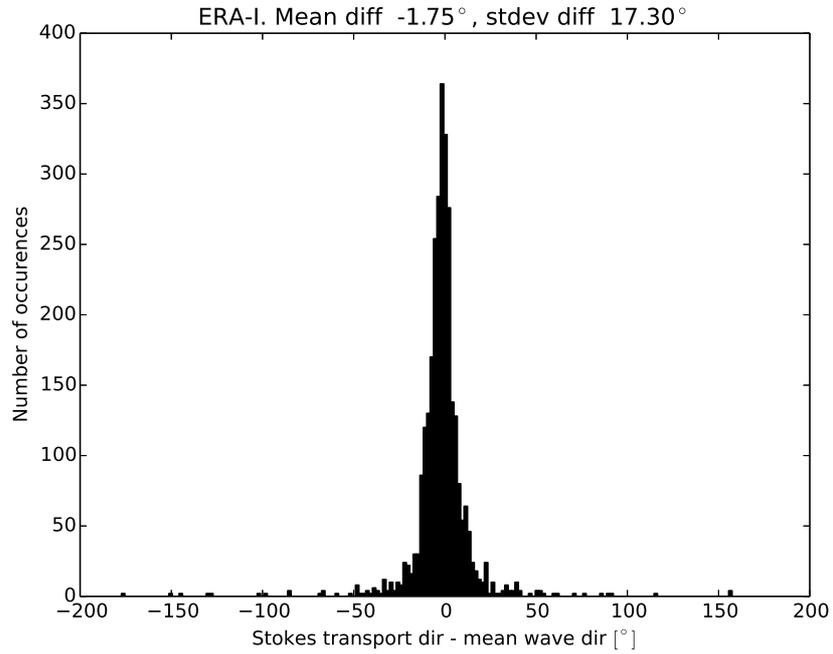}\\
(b)\includegraphics[scale=0.6]{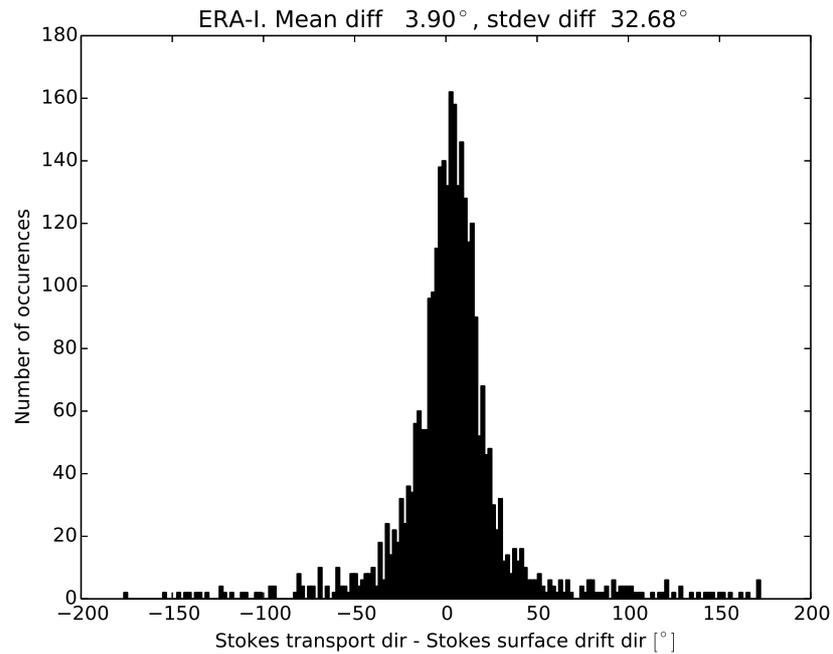}\\
\caption{Panel a: The directional deviation between the Stokes transport and
the mean wave direction (MWD). The average deviation is about $2^\circ$ and
75\% of the time the difference is less than $10^\circ$. 
Panel b: The directional deviation between the Stokes transport and the
surface Stokes drift velocity is larger due to the $f^3$ weighting of the
wave spectrum which gives larger weight to high-frequency wave components.}
\label{fig:dirdiff}
\end{center}
\end{figure}

The mean wave direction (MWD) measured clockwise from north in the direction
the waves are propagating to is defined as
\begin{equation}
  \overline{\theta} = \mathrm{arctan}\left(
  \frac{\int_0^{2\pi}\!\int_0^{\infty} \sin \theta F(f,\theta) \, \dr f \dr\theta}
       {\int_0^{2\pi}\!\int_0^{\infty} \cos \theta F(f,\theta) \, \dr f \dr\theta}
  \right).
  \label{eq:MWD}
\end{equation}
It is of interest to assess how well it approximates the direction of the
Stokes transport since it is a standard output parameter of many wave models
\citep{wam40r1} whereas the Stokes transport is generally not. Panel a of
\Fig{dirdiff} shows the deviation of the Stokes transport from the MWD in the
model location in the north Atlantic during 2010. The average deviation is
about $2^\circ$ and 75\% of the time the difference is less than $10^\circ$. In
contrast, Panel b shows a much larger deviation between the direction of the
Stokes transport and the surface Stokes drift velocity. This is due to the
sensitivity to high-frequency wave components arising from the third power
of the frequency $f$ under the integral in \Eq{uvfth}. It will therefore
in general be better to estimate the transport direction from the mean wave
direction rather than from the surface Stokes direction.

\section{Stokes Profiles under Measured Spectra in the North Sea}
\label{sec:ekofisk}
\begin{figure}[h]
\begin{center}
(a) \includegraphics[scale=0.6]{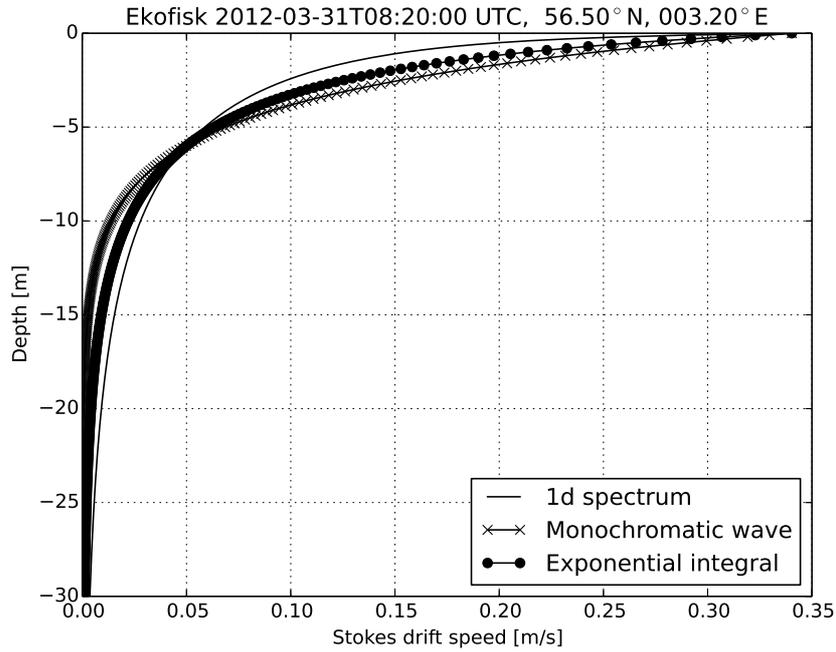}\\
(b) \includegraphics[scale=0.6]{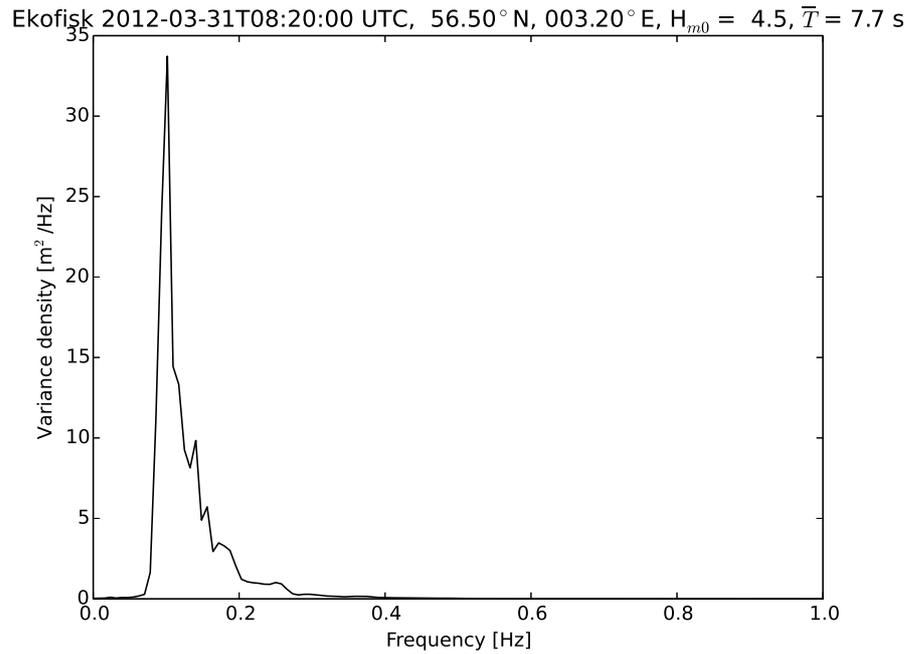}\\
\caption{Panel a: The Stokes drift profile under the one-dimensional spectrum
at Ekofisk in the central North Sea. A better fit is found with the exponential
integral profile even in the presence of high-frequency spectral noise.
Panel b: The spectrum was computed 
from a 20-min 2 Hz time series from a Datawell Waverider buoy. The spectrum is
plotted up to the Nyquist frequency at 1 Hz. High-frequency noise affects the
surface Stokes drift estimates somewhat.}
\label{fig:obs}
\end{center}
\end{figure}

\begin{figure}[h]
\vspace{4mm}
\begin{center}
\hspace{5mm}
\includegraphics[scale=0.9]{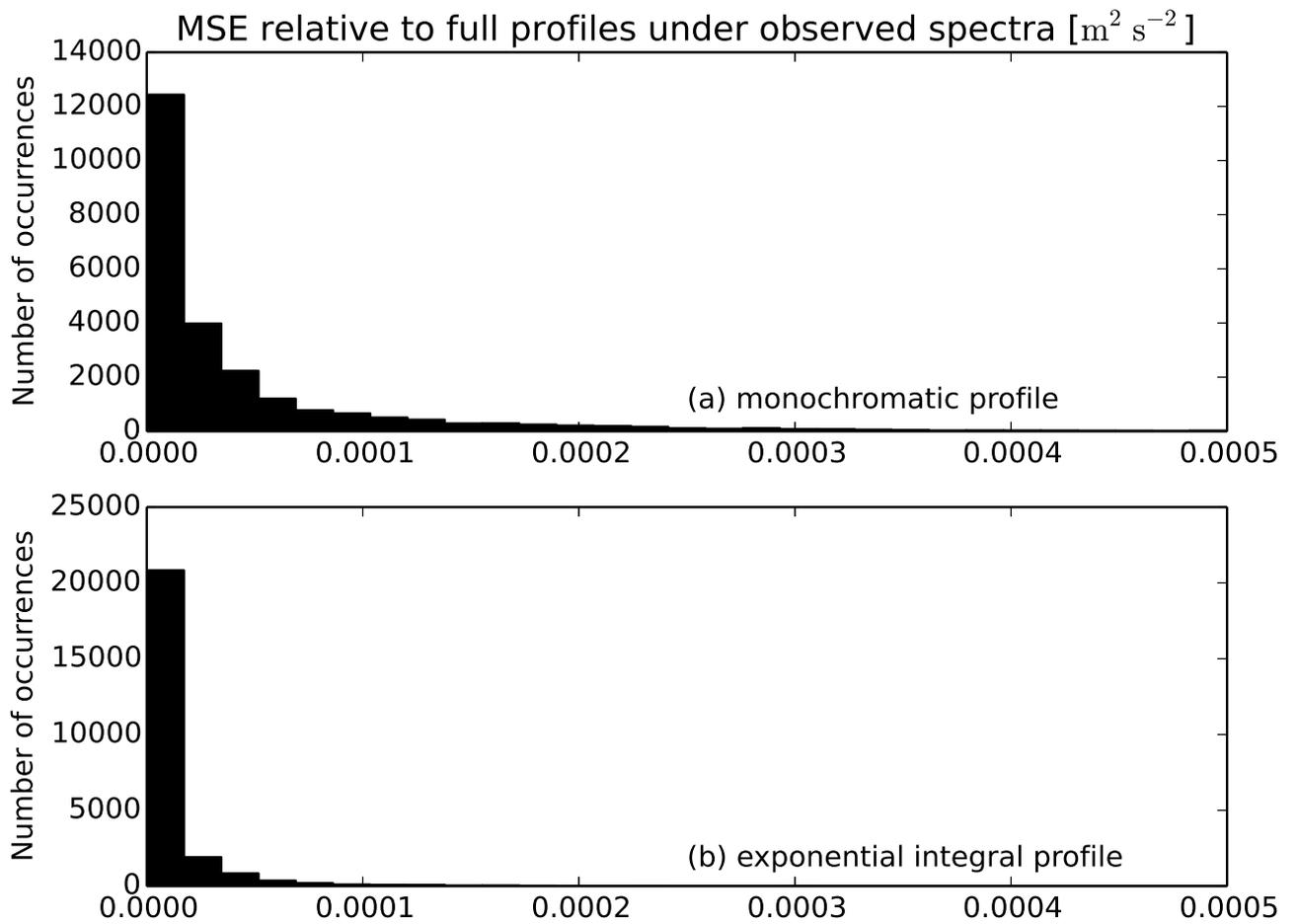}\\
\vspace{4mm}
\caption{Panel a: The MSE between the full Stokes
profile computed under a 20-min 2 Hz time series from a Datawell Waverider buoy
at Ekofisk and the monochromatic profile to 30 m depth (vertical resolution
0.1 m).  Panel b: The MSE of the exponential integral profile is
on average about 40\% that of the monochromatic profile shown in Panel a.}
\label{fig:rmsobs}
\end{center}
\end{figure}

A directional Datawell Waverider buoy anchored near Ekofisk in the
central North Sea (56.5$^\circ$N, 003.2$^\circ$E) provided one year
of data (2012) at 2 Hz sampling rate (location marked with asterisk in
\Fig{mwd_hist_and_location}b).  24,894 spectra from 20-min time series were
computed, with some gaps (about 5\% of the time series were either missing
or discarded). The 2,400 measurements in the 20-min time series were split
into 8 non-overlapping parts and a Hann window [\citet{pre07} pp 656--660
and \citet{christensen13}]  was applied to each chunk,
\begin{equation}
   w_n = \frac{1}{2}\left(1-\cos \frac{\pi n}{N}\right).
   \label{eq:hann}
\end{equation}
Here the taper width $N$ was set to 32.
Finally, the power spectrum was smoothed with a triangular filter
\begin{equation}
   \overline{F}_j = (F_{j-1} + 2F_j + F_{j+1})/4.
   \label{eq:psmooth}
\end{equation}
The results are very similar to what is found for the modeled spectra
(\Fig{obs}), with an MSE for the exponential integral profile 60\%
lower than for the monochromatic profile (cf \Fig{rmsobs}). The high-frequency
part of the observed spectra tends to be rather noisy. This affects the surface
Stokes drift and makes it sensitive to the high-frequency tail contribution
(\ref{eq:utail})-(\ref{eq:uhf}), clearly illustrated by the spectrum shown in
\Fig{obs}b. Nevertheless, the match is better both with an without the added
tail (about 50\% reduction in MSE with tail added, not shown).  Although the
water depth at Ekofisk is only 70~m, the deep water approximation will hold
in most cases and shallow water effects for the highest storm situations
are not likely to affect the results significantly.

\section{Recommendations for Approximate Stokes Drift Profiles}
\label{sec:conclusion}
The alternative profile proposed here has been shown to be a
better approximation than the monochromatic approximation for both
theoretical spectra, modeled 2-D spectra in the open ocean and 1-D observed
spectra. Utilizing this alternative profile comes at no added cost since the
computation relies on the same two parameters required for the monochromatic
profile, namely the Stokes transport, $\mathbf{V}_\mathrm{s}$, and the
surface Stokes drift velocity, $\mathbf{v}_0$. We also found that in the
open ocean the mean wave direction serves as a good proxy for the Stokes
transport direction. It is a significantly better substitute than the surface
Stokes drift direction.  Furthermore, the one-dimensional first order moment,
$m_1$, is found to correlate well with the magnitude of the two-dimensional
transport, $|\mathbf{V}_\mathrm{s}|$.  A reduction factor of 0.86
is appropriate in open ocean conditions.

Discretized spectra add a diagnostic high-frequency tail, see \Eq{Fhf}. We find
that adding the contribution from the tail gives an important contribution
to the Stokes drift velocity in the upper half meter in the open ocean.
Its impact rapidly decays, and below 0.5 m the difference is marginal (Panel
a, \Fig{uhf_profile}).  This has implications for the computation of the
gradient of the Stokes drift in the uppermost part of the ocean. Neither of
the approximate profiles match the gradient in the upper half meter well,
and this is important to keep in mind for future studies of upper-ocean
turbulence.  We note again that although it is numerically inexpensive to
treat the high-frequency contribution to the profile separately, unless it
has been explicitly archived, its reliance on the full 2-D spectrum makes this
approach impractical for many applications where the spectrum is not available.

We conclude that the proposed Stokes drift profile is a much closer match than
the commonly used monochromatic profile both in terms of speed and shear.
Although neither profile is a good match for the shear in the upper half meter,
even here the new profile offers a slight improvement over the monochromatic
profile. As Langmuir turbulence depends sensitively on the Stokes drift
shear the question of whether approximate profiles can be found that more
closely mimic the gradient in the uppermost half meter merits further work.

\section*{Acknowledgment} 
This work has been carried out with support from the European Union FP7
project MyWave (grant no 284455). Many thanks to Magnar Reistad at the
Norwegian Meteorological Institute for providing the buoy data. Thanks also
to the two reviewers who made the paper a much better one.

{\clearpage}
\bibliography{./database,./references} 

\end{document}